\documentclass[11pt]{article}
\usepackage{hyperref}
\pdfoutput=1

\begin{document}

\title{Infrared video of a warm water surface in the presence and
absence of surfactant monolayers}

\author{S. M. Bower \& J. R. Saylor \\ Department of Mechanical Engineering \\
  Clemson University, Clemson, SC 29634, USA}

\maketitle

\begin{abstract}
Infrared (IR) videos are presented which show a warm water surface
undergoing convective processes.  These fluid dynamics videos show the
water surface with: 1) no surfactant monolayer material present, 2) a
liquid-phase monolayer of oleyl alcohol, and 3) a solid-phase
monolayer of cetyl alcohol.
\end{abstract}

\section{Introduction}

This
\href{http://ecommons.library.cornell.edu/bitstream/1813/14124/2/DFDV2.mpg}{video}
(\href{http://ecommons.library.cornell.edu/bitstream/1813/14124/3/DFDVideoLowQualV1.mpeg}{low
quality version}) provides IR visualizations of a warm water surface
(T $\approx$ 45$^{\circ}$C) under several air/water-interfacial
conditions: namely with a clean surface (no surfactant material
present), a liquid-phase monolayer present, and a solid-phase
monolayer present. The clean water surface exhibits fine-scale
structures and an average surface temperature that is slightly less
than the bulk water temperature.

When the liquid-phase monolayer of oleyl alcohol is spread across the
surface, it imposes a constant-elasticity boundary condition. The
average surface temperature is noticeably less than that of the clean
surface and the fine-scale structures vanish when the liquid-phase
monolayer is present.

A solid-phase monolayer of cetyl alcohol is then compared to the oleyl
alcohol surfactant by depositing talc powder onto both
monolayer-contaminated water surfaces. Because the solid-phase
surfactant imposes a nearly rigid boundary condition at the interface,
the monolayer both supports the talc powder at the surface and is not
deformed by the subsurface motion of the water bulk.

Lastly, air is blown across both the liquid-phase and solid-phase
surfactants. While wind shear cools and deforms the surface with the
liquid-phase monolayer, the solid-phase monolayer tends to resist
surface motion and does not cool noticeably. When considering that
these monolayers are only one molecule thick, these visualizations are
impressive and reveal the significance that these monolayers have on
transport processes.

\end{document}